\documentstyle[aps,eqsecnum,preprint,floats,epsf,psfig]{revtex}

\begin{document}
\draft

\title{
\Large\bf Nonperturbative Determination \\ of Heavy Meson Bound States}

\author{{\bf Chi-Yee Cheung$^a$ and Wei-Min Zhang$^b$} \\
$^a$Institute of Physics, Academia Sinica, Taipei, Taiwan 115, R.O.C.\\
$^b$Department of Physics, National Cheng-Kung University, Tainan, 
Taiwan 701, R.O.C.}

\date{Dec. 20, 1998}

\maketitle

\begin{abstract}
In this paper we obtain a heavy meson bound state equation  
from the heavy quark equation of motion in heavy quark effective theory 
(HQET) and the heavy meson effective field theory we developed very 
recently. The bound state equation is a covariant extention of the
light-front bound state equation for heavy mesons derived from
light-front QCD and HQET.
We determine the covariant heavy meson wave function   
variationally by minimizing the binding energy $\overline{\Lambda}$.  
Subsequently the other basic HQET parameters $\lambda_1$ and
$\lambda_2$, and the heavy quark masses $m_b$ and $m_c$ can also be 
consistently determined. 
\end{abstract}

\vspace{0.5in}


\newpage

\section{Introduction}
 
Very recently, we have developed
an effective field theory to describe the nonperturbative 
QCD dynamics of heavy mesons \cite{cheung98,cheng98}. 
This theory has incorporated in it heavy quark effective theory 
(HQET), and is thus consistent with heavy quark symmetry (HQS) in the 
infinite quark mass ($m_Q\rightarrow\infty$) limit.
In this effective field theory, a heavy meson is considered as a composite 
particle consisting of a reduced heavy quark (heavy quark in the  
$m_Q\rightarrow\infty$ limit) coupled with the light degrees of freedom.   
A structure function $\Psi(v \cdot p_q)$,  
corresponding to the wave function of the heavy meson bound state, 
is introduced to describe the 
quark-meson vertex in the effective Lagrangian.  By combining such
an effective Lagrangian for the composite heavy mesons with the 
$1/m_Q$ expansion of the heavy quark QCD Lagrangian, we can 
systematically evaluate various $1/m_Q$ corrections 
to heavy meson properties in a covariant field-theoretic framework 
\cite{cheng98}.  

In our previous works \cite{cheung98,cheng98}, the 
heavy meson wave function $\Psi(v \cdot p_q)$ is taken to be a 
phenomenological function.  Specifically it is assumed to be 
a Lorentzian function, $\Psi(v \cdot p_q) 
= 1/(v\cdot p_q +\omega)^n$, in which the 
parameters $\omega$ and $n$ are determined phenomenologically by fitting to 
the $B$ meson decay constant $f_B$; hence its relation to  
nonperturbative low-energy QCD dynamics is not clear. 
In this paper, we attempt to determine the heavy meson
wave function from a QCD-based bound state equation.  The bound  
state equation of heavy mesons is obtain from the heavy quark 
equation of motion in HQET which is incorporated in our effective field 
theory.  This heavy meson bound state equation is a
covariant extension of the light-front bound state equation we 
obtained recently from light-front QCD and HQET\cite{zhang97}. 
With the bound state equation for $\Psi(v \cdot p_q)$, we can use 
a variational procedure to determine its soluton.  
 
The paper is organized as follows: In Sec.~II, we derive from 
the heavy quark equation of motion in the heavy quark limit and our 
effective heavy meson field theory the heavy meson bound state equation.
Then a variational method is used to determine the heavy meson wave function
$\Psi(v \cdot p_q)$ by minimizing the reduced mass (binding energy) of the 
heavy meson $\overline{\Lambda}$.  
In the literature, $\overline{\Lambda}$ has  
only been estimated either from heavy meson mass spectrum or from inclusive 
heavy meson decay data \cite{lamb}.  A direct calculation of 
$\overline{\Lambda}$ has not been carried out to date.  In Sec.~III, we  
consistently calculate the other two basic HQET parameters $\lambda_1$ 
and $\lambda_2$ directly from the 
operators ${\cal O}_1$ and ${\cal O}_2$ in HQET.  Having   
self-consistently calculated $\lambda_1$ and $\lambda_2$, 
we can then extract from experimental heavy 
meson masses the heavy quark masses ($m_b$, $m_c$) at the low energy 
scale where our effective field theory applies.  Such a procedure 
establishes a link between our heavy meson effective field theory  
and nonperturbative QCD dynamics.  Finally a summary is given in Sec.~IV.

\section{Heavy meson bound state equation in effective field 
theory}

\subsection{Heavy meson effective field theory}

Before deriving the heavy meson bound state equation, we shall
briefly review the heavy meson effecive field theory we developed
recently \cite{cheung98,cheng98}. 
In heavy quark effective theory (HQET) \cite{Georgi90}, the QCD 
Lagrangian is expanded in powers of $1/m_Q$:
\begin{eqnarray}
	{\cal L}_{QCD} &=& \overline{h}_v i v \cdot D h_v + \overline{q}
		(i \! \not \! \! D - m_q) q - {1\over 4} F_a^{\mu \nu}
		F_{a\mu \nu}  \nonumber \\ 
	& & + \sum_{n=1}^\infty \Bigg({1 \over 2m_Q} \Bigg)^n 
		\overline{h}_v (i \! \not \! \! D_\bot)(-iv \cdot D)^{n-1}
		(i \! \not \! \! D_\bot) h_v    \nonumber \\
	&=& {\cal L}_0 + {\cal L}_{m_Q}	\,  \label{1me}
\end{eqnarray}
where the reduced heavy quark field $h_v$ is defined from the full 
heavy quark field $Q$ by
\begin{equation}
	Q(x) = e^{-im_Q v \cdot x} h_v + O(1/m_Q) ~~, ~~~
		{1 + \not \! v \over 2} h_v = h_v  \, ,
\end{equation}
$v^\mu$ is the heavy quark four velocity which is defined to be equal 
to that of the parent heavy meson, 
and $D^\mu_\bot \equiv D^\mu - v^\mu v \cdot D$.  
It is easily seen that the leading term 
${\cal L}_0=\bar h_v i v\cdot D h_v$ in Eq.(\ref{1me})
possesses a $SU_f(N_f)\otimes SU_s(2)$ flavor-spin symmetry 
(or simply heavy quark symmetry, HQS), whereas the symmetry breaking 
part ${\cal L}_{m_Q}$ contains $1/m_Q$ corrections to ${\cal L}_0$.

The effective Lagrangian we have constructed to describe the low energy 
dynamics of heavy mesons reads \cite{cheung98,cheng98}:
\begin{eqnarray}
	{\cal L}_{\rm eff} 
	  &=& \overline{h}_v i v \cdot \partial h_v + \overline{q}(i \! 
		\not \! \! D - m_q) q 
		-{1\over4} F^{\mu\nu}_a F_{a\mu\nu} \nonumber \\
	  & & + P^\dagger_{v} \Big(iv \cdot \stackrel{\leftrightarrow}
		{\partial} - 2\overline{\Lambda} \Big) P_{v}  
	     - V^{\mu\dagger}_{v} \Big(iv \cdot \stackrel{\leftrightarrow}
		{\partial} - 2\overline{\Lambda} \Big) V_{v\mu}  \nonumber \\
	  & & - \Big(\overline{h}_v i\gamma_5 q_v P_{v} - 
		\overline{h}_v \gamma_\mu q_v V_{v}^\mu  + h.c. \Big) 
			\label{elcy1}  \nonumber \\
	  & & + \sum_{n=1}^\infty \Bigg({1 \over 2m_Q} \Bigg)^n 
		\overline{h}_v (i \! \not \! \! D_\bot)(-iv \cdot D)^{n-1}
		(i \! \not \! \! D_\bot) h_v \nonumber \\  
	   &=& {\cal L}^{QqH}_{\rm eff} + {\cal L}_{m_Q} \, ,  \label{hqel}
\end{eqnarray}
where $P_v$ and $V^\mu_v$ represent respectively the composite pseudoscalar 
and vector heavy meson fields which appear only as external states, 
\begin{equation}	\label{lmb}
	\overline{\Lambda} \equiv \lim_{m_Q\rightarrow \infty} (M_H-m_Q) 
\end{equation}
is their common residual mass in the heavy quark limit,
\begin{equation}
	q_v = G\Psi(v \cdot p_q) q
\end{equation}
represents collectively the light degrees of freedom in a heavy meson, where 
$\Psi(v \cdot p)$ is heavy meson wave function mentioned earlier, 
$p_q$ is the light antiquark momentum, and
$G$ is the normalization constant given by 
\begin{equation}
	G^{-2} = i \int {d^4 p_q \over (2\pi)^4} {|\Psi(v \cdot p_q)|^2
		\over (\overline{\Lambda} - v \cdot p_q + i \epsilon)^2} 
		\,{ v \cdot p_q + m_q \over (p^2 - m_q^2 + i \epsilon)}  
		\label{wfrn} \, . 
\end{equation}

In Eq.(\ref{hqel}), ${\cal L}^{QqH}_{\rm eff}$ 
obeys HQS and defines the nonperturbative 
effective coupling of heavy mesons with their constituent heavy and light 
quarks, while ${\cal L}_{m_Q}$, which contains all the $1/m_Q$ corrections,  
is considered as a perturbation to ${\cal L}^{QqH}_{\rm eff}$. 
A practical calculational scheme has been developed 
in terms of the Feynman diagrammatic rules summarized below 
\cite{cheung98,cheng98}: 

(i) A heavy meson bound state ($p_H=\overline{\Lambda}v$) couples to its 
constituent heavy and light quarks via a three-point vertex
\begin{equation}
\begin{picture}(65,30)(0,38)
\put(0,37){$- - -$}
\put(33,40){\circle*{6} }
\put(33,41){\line(2,1){20}}
\put(33,40){\line(2,1){20}}
\put(33,39){\line(2,-1){20}}
\end{picture}
	: ~~~~~~~ -i G  \Psi (v \cdot p_q) \Gamma_H \, ,
\end{equation}
\begin{equation}
\begin{picture}(65,30)(0,38)
\put(20,37){$- - -$}
\put(20,40){\circle*{6} }
\put(20,41){\line(-2,1){20}}
\put(20,40){\line(-2,1){20}}
\put(20,39){\line(-2,-1){20}}
\end{picture}
	: ~~~~~~~ -i G  \Psi^* (v \cdot p_q) \Gamma_H \, ,
\end{equation}
where $\Gamma_P = i \gamma_5$ for pseudoscalar
$(H=P)$, $ \Gamma_V = - \! \not \! \epsilon$ for vector mesons 
$(H=V)$, and $p_q$ is the momentum of the light degrees of freedom.
Each vertex also carries a momentum conservation factor of 
$ (2\pi)^4 \delta^4 (\overline{\Lambda}v-k-p_q)$. 
 
(ii) The internal heavy quark and light quark propagators are, 
\begin{eqnarray}
\begin{picture}(65,30)(0,38)
\put(0,40.5){\line(1,0){40}}
\put(19,40){\vector(1,0){2}}
\put(0,39.5){\line(1,0){40}}	
\put(19,28){$k$}
\end{picture}
	&:& ~~~~~~ i{\not \! v + 1 \over 2 (v \cdot k + i\epsilon)} \, ,\\ 
\begin{picture}(65,30)(0,38)
\put(0,40){\line(1,0){40}}
\put(21,40){\vector(-1,0){2}}	
\put(17.5,30){$-p_q$}
\end{picture}
 &:& ~~~~~~ i { - \!\not \! p_q + m_q \over p_q^2 - m_q^2 + i \epsilon} 
\, , \end{eqnarray} 
respectively, where $k$ is the residual momentum of the heavy quark,
and $m_q$ the constituent mass of the light quark. 

(iii) An integration factor $\int {d^4 p/(2\pi)^4}$ is associated with 
each internal momentum $p$, and a factor $(-1)$ should be included for each 
fermion loop. 

(iv) For all other lines and vertices not attached to heavy mesons 
(mostly from ${\cal L}_{m_Q}$), standard Feynman rules apply.
Hence $1/m_Q$ corrections to various quantities 
can be calculated using standard field theoretic perturbative approach 
\cite{cheng98}. 

\subsection{Heavy meson bound state equation}

In HQET, the reduced heavy quark field $h_v$ 
obeys the following equation of motion
\begin{equation}
	iv\cdot D h_v = 0,
\end{equation}
where $D^\mu=\partial^\mu -ig A^\mu$ is the QCD covariant derivative.

To ensure that our effective field theory, Eq.(\ref{hqel}), is 
consistent with low energy QCD dynamics of HQET, Eq.(\ref{1me}), 
we require that in the heavy quark limit, 
the heavy meson state $|H(v) \rangle$ satisfies 
\begin{equation}	\label{hqem}
	\langle H_v | \overline{h}_v i v \cdot D h_v | H_v \rangle = 0.
\end{equation}
Strickly speaking, this matrix element must be calculated in a 
nonperturbative manner of QCD, which is however beyond our present 
day knowledge of 
QCD.  To circumvent this predicament, we shall adopt an effective 
gluon propagator which contains a 
long-distance contribution as well as the familiar perturbative one
\cite{Munczek}: 
\begin{equation}  \label{egp}
g_s^2 \Pi^{\mu\nu}(q)=-ig^{\mu\nu}\delta_{ab}
\Bigg({g_s^2\over q^2+i\epsilon}+i4\pi^2\Lambda^2_c\delta^4(q)\Bigg)
\end{equation}
where $g_s$ is the QCD coupling constant, and
the coefficient $\Lambda_c^2$ characterizes the confining strength.
Note that the regularized infrared singular term $\delta^4(q)$
can be effectively considered as a regularization of the linearly confining 
interaction $1/q^4$ \cite{Munczek}.  
Representing this effective gluon propagator by a double wavy line, 
we can depict the left hand side of Eq. (\ref{hqem}) pictorially 
by Feynman diagrams shown in Fig. 1.

As discussed in \cite{cheung98,cheng98}, from 
the constraint that it forbids on-mass-shell decay of the heavy meson 
into $Q\bar q$, a possible form of the heavy meson wave function 
$\Psi(v\cdot p_q)$ is given by 
\begin{equation}
	\Psi(v \cdot p_q)= (\overline\Lambda - v \cdot p_q) 
                   \varphi(v \cdot p_q)	\label{Psi},
\end{equation}
where the function $\varphi(v \cdot p_q)$ is regular at  $v \cdot p_q 
= \overline\Lambda$.  
Then, applying the diagrammatic rules stated in Sec.~IIA to Fig. 1, 
we can easily translate Eq.(\ref{hqem}) into the following expression,
\begin{eqnarray}
	i G^2\int {d^4p_q \over (2\pi)^4} &&|\varphi(v\cdot p_q)|^2 
	{v\cdot p_q +m_q \over p^2-m_q^2}(\overline{\Lambda}-v\cdot p_q)
		\nonumber \\
	&& +C_f G^2 \int {d^4p_q \over (2\pi)^4}{d^4p'_q \over(2\pi)^4}
		\varphi(v\cdot p_q)\varphi^*(v\cdot p'_q) \Bigg({4\pi
		\alpha_s \over (p_q-p'_q)^2}+i4\pi^2\Lambda^2_c 
		\delta^4(p_q-p'_q) \Bigg)\nonumber \\
	&& ~~~~~~~~~~~~~~~~ \times {(v\cdot p_q+m_q)(v\cdot p'_q+m_q)
	+v\cdot p_q v\cdot p'_q - p_q\cdot p'_q \over (p_q^2
	-m_q^2)({p'}_q^{2}-m_q^2)}\nonumber \\
	&&=0,	\label{eqm2}
\end{eqnarray}
where the color factor $C_f=(N_c^2-1)/2N_c =4/3$ for $N_c=3$. 
The above equation can be rewritten as an equivalent 
bound state equation for heavy mesons:
\begin{eqnarray}
	 (\overline{\Lambda}-v\cdot p_q)\varphi(v\cdot p_q)&=&
		C_f \int {d^4p'_q \over(2\pi)^4} \Bigg({i 4\pi \alpha_s\over 
		(p_q-p'_q)^2}-4\pi^2 \Lambda^2_c \delta^4(p_q-p'_q)	
 		\Bigg)\nonumber \\
	&& \times {(v\cdot p_q+m_q)(v\cdot p'_q+m_q)
		+v\cdot p_q v \cdot p'_q - p_q \cdot p'_q \over
		(v\cdot p_q +m_q)({p'}^{2}_q-m_q^2)} \varphi(v\cdot p'_q),
		\label{bse}
\end{eqnarray}
which is indeed a covariant generalization of the light-front 
bound state equation for heavy mesons we have previously derived 
\cite{zhang97}, and it is also the heavy quark limit of 
the Bethe-Salpeter equation for heavy mesons.

Using the normalization condition of $\Psi(v\cdot p_q)$,
we can simplified Eq.(\ref{eqm2}) to 
\begin{eqnarray}
	\overline{\Lambda} &=& i G^2 \int {d^4p_q \over (2\pi)^4}
		|\varphi(v\cdot p_q)|^2 
	{v\cdot p_q +m_q \over p^2-m_q^2}~ v\cdot p_q \nonumber \\
	&& - C_f G^2 \int {d^4p_q \over (2\pi)^4}{d^4p'_q \over
		(2\pi)^4}\varphi(v\cdot p_q)\varphi^*(v\cdot p'_q) \Bigg(
		{4\pi\alpha_s \over (p_q-p'_q)^2} 
	+ i4\pi^2\Lambda^2_c 		\delta^4(p_q-p'_q) \Bigg)\nonumber \\
	&& ~~~~~~~~~~~~~ \times {(v\cdot p_q+m_q)(v\cdot p'_q+m_q)
		+ v\cdot p_q v\cdot p'_q - p_q\cdot p'_q \over (p_q^2
		-m_q^2)({p'}_q^{2}-m_q^2)}.	\label{lmb1} 
\end{eqnarray} 
The physical meaning of the above result is clear. 
The heavy meson binding energy $\overline{\Lambda}$ defined by 
Eq.(\ref{lmb}) consists of two parts:  One is the light quark kinetic 
energy contribution $\langle v\cdot p_q\rangle$, 
and the other is the interaction energy of the heavy-light system. 
Note that, due to the effective gluon propagator (\ref{egp}) used, 
the interaction energy contains a confining contribution, as well 
as the color Coulomb contribution.  

\subsection{Variational determination of heavy meson wave function 
$\varphi(v\cdot p_q)$} 

In this section we shall determine the heavy meson wave function 
$\varphi(v\cdot p_q)$ through a variational procedure.
The first step is to choose an appropriate trial wave function. 
As discussed in \cite{cheung98,cheng98}, we demand that the wave 
function should be relatively well behaved, by which we mean that,
when continued into the complex $p_q$ plane, 
(i) $\varphi(v \cdot p_q)$ is  analytic except for isolated 
singularities, and (ii) it vanishes ``fast enough" 
as $|v \cdotp q|\rightarrow\infty$.  
These two constraints also allow us to 
easily derive various relativisitic and/or light-front quark 
models results as special cases in our effective field theory 
\cite{cheung98,cheng98}.  

In various existing light-front or relativistic quark models, 
two types of distribution functions have 
been widely adopted for phenomenological hadronic wave functions.
One is the exponential(Gaussian)-type and the other  
the Lorentzian-type.  The former can be excluded since it does 
not simultaneously satisfy the two constraints stated above 
\cite{cheung98,cheng98}. 
On the other hand, we find that Lorentzian-type functions of the form,
\begin{equation}  \label{lwf}
	\varphi_n (v \cdot p_q) = {1 \over (v \cdot p_q + \omega -
 		i\epsilon)^n} \, ,
\end{equation}
do meet our requirements \cite{cheung98,cheng98}, 
and will be taken as the working trial heavy meson wave 
function in this work.  
It is interesting to note that such a 
choice of $\varphi(v\cdot p_q)$ corresponds to a covariant 
light-front wave function \cite{cheng98a} of the form
\begin{equation}
	\varphi^{n}_{LF} (v \cdot p_q) = {\cal N}_L {\sqrt{ v\cdot p_q
		+ m_q} \over (v \cdot p_q + \omega )^n  } ~~~
		~~ n > 4 \, ,
\end{equation}
where ${\cal N}_L$ is a normalization constant, $p_q^2 = m_q^2$,
and $n > 4$ is required to ensure convergence.

For the sake of demonstration, we first consider the naive case of 
ignoring of the confining interaction between the 
heavy quark and the light degrees of freedom, {\it i.e.} $\Lambda_c=0$,
so that color Coulomb interaction is solely responsible for binding the 
quarks together.  
We find that, for $n=6$, a minimum with $\overline{\Lambda}=0.192$ GeV  
can be found at $\omega=0.03$.  
However a wave function with such a small $\omega$ is unrealistic 
because it contains too little high momentum components.  In other words, 
the wave function is too extended in configuration space, as would be 
expected since we have ignored the confining part of the inter-quark
interaction.  

With an appropriate non-zero ${\Lambda_c}$, 
physically sensible results can be obtained. 
In what follows, we will take $\alpha_s=0.44$ and $m_q=0.22$ GeV, 
which are close to the values used in our previous work 
\cite{cheng98}.  
With the confinement strength taken to be $\Lambda_c=0.665$ GeV, 
we determine the wave function parameter $\omega$ by minimizing 
$\overline{\Lambda}$.  The results are listed in the Table I, where 
\begin{equation} 
	\overline{\Lambda} = E_{\rm ke} + E_{\rm conf} + E_{\rm coul}
		= 0.426 \sim 0.430  ~~{\rm GeV}
\end{equation} 
which is not sensitive to the choice of $n$.  
This value of $\overline{\Lambda}$ is 
somewhat larger than our previous result of 
$\overline{\Lambda}=0.33\sim 0.34$ GeV obtained in \cite{cheng98}. 
The difference is mainly due to the confinement effect which is not 
considered there. 

\newpage 
{\footnotesize 
Table I. The wave function parameter $\omega$ nonperturbatively 
determined from bound state equation. 
\begin{center} 
\begin{tabular}{|c|c|c|c|c|c|}  
\hline\hline $\phi_n (v \cdot p_q)$ & ~$\omega$~ & ~$\langle 
v\cdot p_1\rangle $~ &~$E_{\rm conf}$~ &~ $E_{\rm coul}$~ &~ 
$\overline{\Lambda}$ \\ \hline
$n=6$ & 0.80 & 0.443  & 0.147  & -0.160 & 0.430 \\ \hline 
$n=7$ & 1.01 & 0.440  & 0.147  & -0.158 & 0.427 \\ \hline 
$n=8$ & 1.24 & 0.438  & 0.144  & -0.157 & 0.426 \\ \hline\hline 
\end{tabular}
\end{center}}
\vskip 0.5cm
 
It is worth emphasizing that, in our previous work
\cite{cheng98}, the parameter $\omega$ is fixed 
phenomenologically by fitting to $f_B=180$ MeV from lattice 
gauge calculations (it is experimentally unknown).  
In the present work, we have dynamically determine the wave 
function $\varphi(v \cdot p)$ by minimizing the blinding energy 
$\overline{\Lambda}$ from the QCD-based bound state equation.  
This is a significant improvement which helps to raise the 
predictive power of our heavy meson effective theory.  

\section{Pseudscalar and vector heavy meson masses}

In this section, we shall 
use the bound state solution obtained in the previous section to 
calculate heavy meson masses up to order $1/m_Q$.  
This will in turn yield more consistent results for $\lambda_1$ and 
$\lambda_2$ than those obtained in \cite{cheng98}.

In the heavy quark limit, the heavy meson masses can be written as
$M_H = m_Q + \overline{\Lambda}$.  The correction to $M_H$ in HQET 
comes mainly from the leading HQS-breaking $1/m_Q$ corrections 
[see Eq.~(\ref{hqel})]: 
\begin{equation}
	{\cal L}_1 = \overline{h}_v (i D_\bot)^2 h_v + {g_s\over 2}
		\overline{h}_v \sigma_{\mu \nu}G^{\mu \nu} h_v 
		= {\cal O}_1 + {\cal O}_2 \, ,   \label{1ml}
\end{equation}
where $\sigma_{\mu \nu} = {i\over 2}[\gamma_\mu, \gamma_\nu]$ and
$G^{\mu \nu} = {i \over g_s}[D^\mu, D^\nu]$. 
With these $1/m_Q$ corrections included, the heavy meson masses can be 
expressed as \begin{eqnarray}  \label{massform}
	M_H = m_Q + \overline{\Lambda} - {1\over 2 m_Q}(\lambda_1
		+ d_H \lambda_2 ) \, , \label{mass}
\end{eqnarray} 
where $\lambda_1$ comes from ${\cal O}_1$, $d_H\lambda_2$ from
${\cal O}_2$ (see Fig.~3), and $d_H  = 3 (-1)$ for pseudoscalar (vector) 
mesons \cite{cheng98}. 
In Eq. (\ref{mass}), $\lambda_1$ parametrizes the common mass 
shift, and $\lambda_2$ accounts for the so called hyperfine mass 
splitting between the vector and pseudoscalar heavy meson.  
They can be directly calculated once the heavy meson wave function 
$\varphi(v\cdot p)$ is determined from the variational approach 
presented in Sec.~IIC. 

Expressions for $\lambda_1$ and $\lambda_2$ are essentially the same as 
those given in \cite{cheng98}, except that here we use the effective 
gluon propagator (\ref{egp}),
which is more appropriate for physics in the low energy regime.   
Thus we have 
\begin{eqnarray}
	\lambda_1 &=& i 2 G ^2\int {d^4 p_q\over (2\pi)^4} |\varphi( v\cdot 
		p_q)|^2 {v \cdot p_q + m_q \over p_q^2 - m_q^2 }
		~\Big[p_q^2 - (v \cdot p_q)^2\Big] \nonumber \\
	& & - 2C_fG ^2 \int {d^4 p_q \over (2\pi)^4} {d^4 p'_q 
		\over (2\pi)^4} \varphi(v \cdot p_q)\varphi^*( v \cdot p'_q)
		{ 1 \over (p_q^2 - m_q^2)({p'}_q^2 - m_q^2)}  \nonumber \\ 
	& & ~~~~~~~~~~~~~~~~~ \times \Bigg({ 4\pi \alpha_s \over 
		(p_q-p'_q)^2} + i4\pi^2 \Lambda^2_c \delta^4(p_q-p'_q)
		\Bigg)  \nonumber \\
	& & ~~~~~~~~~~~~~~~~~ \times \Big\{(v \cdot p_q + m_q)[(p_q+p'_q)
		\cdot p'_q - v \cdot (p_q+p'_q) v \cdot p'_q]
		\nonumber \\
	& & ~~~~~~~~~~~~~~~~~~~~~~+ (v \cdot p'_q + m_q)[(p_q+p'_q) \cdot 
		p_q - v \cdot (p_q+p'_q) v \cdot p_q] \Big\} \, ,
		\label{lam1}
\end{eqnarray}
and
\begin{eqnarray}
	\lambda_2 &=& {4\over 3} C_fG ^2 \int {d^4 p_q \over (2\pi)^4}
 		{d^4 p'_q \over (2\pi)^4} \varphi(v \cdot p_q)
		\varphi^*( v \cdot p'_q){ 1 \over (p_q^2 - m_q^2)
		({p'}_q^2 - m_q^2)}  \nonumber \\ 
	& & ~~~~~~~~~~~~~~~~~ \times \Bigg({ 4\pi \alpha_s \over 
		(p_q-p'_q)^2} + i4\pi^2
		\Lambda^2_c \delta^4(p_q-p'_q)\Bigg)  \nonumber \\
	& & ~~~~~~~~~~~~~~~~~~\times \Big\{(v \cdot p_q + m_q)[(p_q-p'_q) 
		\cdot p'_q - v \cdot (p_q-p'_q) v \cdot p'_q] 
		\nonumber \\
	& & ~~~~~~~~~~~~~~~~~~~~~~- (v \cdot p'_q + m_q)[(p'_q-p'_q) \cdot 
		p_q - v \cdot (p_q-p'_q) v \cdot p_q] \Big\} \, .
		\label{lam2}
\end{eqnarray}
Note that since the chromo-magnetic interaction vanishes at $q=0$ 
($p_q=p'_q$), $\lambda_2$ is actually not affected by the confinement 
part of the gluon propagator.  
Now, we can use a self consistent method to determine uniquely the 
parameters of our effective theory ($\Lambda_c, \omega$), and also those of 
HQET ($\overline\Lambda$, $\lambda_1$, $\lambda_2$, and $m_Q$).  
The procedure goes as follows.  For a given $\Lambda_c$, we 
first determine the wave function parameter $\omega$ by minimizing 
the binding energy $\overline{\Lambda}$ via Eq.(\ref{lmb1}). Then 
we can evaluate $\lambda_1$ and $\lambda_2$ using Eqs.(\ref{lam1})
and (\ref{lam2}).  From Eq.~(\ref{massform}), the hyperfine 
mass splitting is given by 
\begin{equation}	\label{msp}
	\Delta M_{VP} = M_V - M_P = {2 \lambda_2 \over m_Q}  \, ,
\end{equation}
which, together with the experimental data 
$\Delta M_{B^*B}=45.7 \pm 0.4$ MeV and $\Delta 
M_{D^*D}=142.12 \pm 0.07$ MeV \cite{PDG}, uniquely determines 
the heavy quark masses $m_b$ and $m_c$. 
Finally, knowing $M_B =5.279$ GeV, we can recalculate   
$\overline{\Lambda}$ from Eq.(\ref{massform}). 
This resulting $\overline{\Lambda}$ should be equal to that determined 
earlier from variation.  If not, the process is repeated with a new 
$\Lambda_c$ until we reach  consistency.  
Through this self consistent procedure, we can for the
first time determine uniquely all the basic HQET parameters $m_b, 
m_c, \overline{\Lambda}, \lambda_1$ and $\lambda_2$ from QCD-based
dynamics in the low energy regime.  
The final results are listed in the Table II. 

\vskip 0.5cm
{\footnotesize
Table II.  The wave function parameter $\omega$ is variationally
determined by minimizing $\overline{\Lambda}$, and all the other HQET 
parameters are calculated self consistently in the heavy meson effective 
theory with $m_q=0.22$ GeV and $\alpha_s=0.44$.
$\Lambda_c, \omega, m_b, m_c$, and $\overline{\Lambda}$ are in GeV, and 
$\lambda_1$ and $\lambda_2$ in GeV$^2$.  
\begin{center}
\begin{tabular}{|c|c|c|c|c|c|c|c|c|}  
\hline\hline $\Psi_n (v \cdot p_q)$  & ~$\Lambda_c $ & ~$\omega~$ 
&~$\overline{\Lambda}$
& $-{\lambda}_1$~ & ${\lambda}_2$ ~ & ~$\Delta M_{B^*B}$~&~$m_b $~ & $m_c $  \\ \hline
$n=6$ & 0.665 & 0.805 & 0.430 & 0.228 & 0.1116 & 0.0459 & 4.860 & 1.570  \\ \hline
$n=7$ & 0.685 & 1.070 & 0.436 & 0.247 & 0.1114 & 0.0459 & 4.852 & 1.568  \\ \hline
$n=8$ & 0.705 & 1.336 & 0.443 & 0.264 & 0.1112 & 0.0459 & 4.845 & 1.556  \\ \hline
exp. & --- & --- & $ 0.39 \pm 0.11 $\cite{Wise1} & $0.19 \pm 0.1$ \cite{Wise1} & 0.1117$^*$ & $0.0457 \pm 0.004$ & $4.89 \pm 0.05^{**}$ & 
$1.59 \pm 0.02^{**}$  \\ \hline\hline 
\end{tabular}
\end{center}
$^*$In the literature, the so-called ``experimental" value $\lambda_2 
\simeq 0.12$ GeV$^2$ is obtained by assuming $m_b=M_B$.  
One would get $\lambda_2\simeq 0.1117$ GeV$^2$ if the pole-mass $m_b^{\rm 
pole}=4.89$ GeV is used in Eq.~(\ref{msp}) for $\Delta M_{B^*B}=45.7$ MeV. 
\newline$^{**}$ There are no experimental data for the constituent quark 
masses $m_b$ and $m_c$, and the pole masses are displaced \cite{mass}. }

\vskip 0.5cm

\section{Summary and Conclusion}

We know that HQS and confinement mechanism 
are crucial ingredients of nonperturbative QCD dynamics in heavy 
hadrons.  However they are rarely incorporated in quark models in a 
fully covariant manner. 
In this work, we attempt to establish a link between quark model and 
low-energy QCD dynamics in an effective field theoretic approach, in 
which the heavy meson wave function is determined variationally.  Our 
effective theory preserves the simplicity of a conventional quark model; 
in fact, for heavy quark transitions, we can reproduce the 
covariant light-front quark model \cite{cheng98a} as a 
special case.  The light-front quark model has been widely used to describe 
heavy meson structures.  However, because its relation to 
QCD is not clear, it is incapable of describing the hadronic mass spectrum.
Moreover, the model is not fully covariant, so that it can only describe  
physical processes with space-like momentum transfers  
\cite{cheung97,cheng97}. 
In the present formalism, we can study hadronic bound state 
physics in a covariant framework; moreover, a connection to low-energy QCD 
dynamics has also been established.  Thus our heavy meson effective theory, 
together with the nonperturbatively determinated heavy meson wave 
function, provides a useful QCD based framework in which to study heavy 
quark physics in the low energy regime.

We have adopted a self-consistent procedure to determine the heavy 
meson wave function, the binding energy $\overline{\Lambda}$, 
as well as its $1/m_Q$ corrections $\lambda_1$ and $\lambda_2$.
>From these HQET parameters, we can determine the heavy quark 
masses $m_b$ and $m_c$ in the low energy regime by fitting to the 
$B^*-B$ and $D^*-D$ hyperfine mass differences.  
For these static properties,  
our results are consistent with other studies on inclusive
B-meson decays \cite{lamb}.

A great advantage of our formalism is that we can unambiguously study 
hadronic transition processes at arbitrary momentum transfers.  
Using the heavy meson wave function determined in this paper, we
can further calculate dynamically the heavy meson decay constant 
(which is usually considered as an input in various quark 
models for heavy mesons) and other hadronic properties such as the  
Isgur-wise function, $DD^*\pi$ and $DD^*\gamma$ coupling constants,
and also their $1/m_Q$ corrections.  The results will be presented in 
forthcoming papers.
 
\acknowledgements

This work is supported in part by National Cheng-Kung University, 
and National Science Council of the Republic of China under Grants
NSC87-2112-M-001-002, NSC88-2112-M-001-012.

\newpage
\begin{center} {\bf Figure Captions}  \end{center}
\begin{description}
\item[Fig.~1] {Diagram for heavy meson bound states:  (a) 
	kinetic energy contribution and (b) effective one-gluon
exchange contribution which contains the Columbus interaction
plus confining interaction in heavy quark symmetry.}
\end{description}

\end{document}